\documentstyle[12pt,epsfig]{article}
\newlength{\dinwidth}
\newlength{\dinmargin}
\newcommand{\ba}{\begin{array}}
\newcommand{\ea}{\end{array}}
\newcommand{\be}{\begin{equation}}
\newcommand{\ee}{\end{equation}}
\newcommand{\bea}{\begin{eqnarray}}
\newcommand{\eea}{\end{eqnarray}}
\newcommand{\gsim}{\mathrel{\mathop{\kern 0pt \rlap
  {\raise.2ex\hbox{$>$}}} \lower.9ex\hbox{\kern-.190em $\sim$}}}

\def\nn{\nonumber}
\def\cO{{\cal O}}

\def\btheta{{\bar{\theta}}}

\def\vx{{\vec{x}}}
\def\cO{{\cal{O}}}

\def\log{{\rm{log}}}

\def\br{{\overline r}}
\def\brh{{\overline \rho}}
\def\btheta{{\overline \theta}}
\def\bm{{\overline m}}

\def\bl{{\overline \ell}}
\def\bt{{\overline t}}
\def\cO{{\cal O}}
\def\bd{{\overline d}}
\def\vx{{\bf x}}
\def\vy{{\bf y}}

\def\hR{{\hat R}}

\def\be{\begin{equation}}
\def\ee{\end{equation}}
\def\bea{\begin{eqnarray}}
\def\eea{\end{eqnarray}}

\begin{document}
\thispagestyle{empty}
\addtocounter{page}{-1}
\vskip-0.35cm
\begin{flushright}
SNUST-02-0302\\
UK/02-04 \\
TIFR-TH/02-09\\
{\tt hep-th/0203164}
\end{flushright}
\vspace*{0.2cm}
\centerline{\Large \bf Penrose limit, Spontaneous Symmetry Breaking}
\vskip0.3cm
\centerline{\Large \bf and}
\vskip0.3cm
\centerline{\Large \bf Holography in PP-Wave Background~\footnote{
Work supported in part by BK-21 Initiative in Physics
(SNU-Project 2), KOSEF Interdiscplinary Research Grant 98-07-02-07-01-5,
and KOSEF Leading Scientist Program.}
}

\vspace*{1.0cm} 
\centerline{\bf Sumit R. Das${}^{a,b,e}$, Cesar Gomez$^{c,e}$ 
{\rm and} Soo-Jong Rey${}^{d,e}$}
\vspace*{0.7cm}
\centerline{\it Department of Physics and Astronomy,}
\vspace*{0.2cm}
\centerline{\it University of Kentucky, Lexington, KY 40506 \rm USA ${}^a$}
\vspace*{0.35cm}
\centerline{\it Tata Institute of Fundamental Research}
\vspace*{0.2cm}
\centerline{\it Homi Bhabha Road, Mumbai 400 005 \rm INDIA ${}^b$}
\vspace*{0.35cm}
\centerline{\it Instituto de Fisica Teorica, C-XVI Universidad Autonoma,}
\vspace*{0.2cm}
\centerline{\it E-28049 Madrid \rm SPAIN ${}^c$}
\vspace*{0.35cm}
\centerline{\it 
School of Physics \& Center for Theoretical Physics}\vspace*{0.2cm}
\centerline{\it Seoul National University, Seoul 151-742 \rm KOREA
  ${}^d$}
\vspace*{0.35cm}
\centerline{\it
Isaac Newton Institute for Mathematical Sciences, Cambridge CB3 0EH
\rm U.K. ${}^e$}
\vspace*{1cm}
\centerline{\tt das@theory.tifr.res.in 
\hskip0.7cm Cesar.Gomez@uam.es
\hskip0.7cm sjrey@gravity.snu.ac.kr
}

\vspace*{0.8cm}
\centerline{\bf abstract}
\vspace*{0.3cm}
We argue that the gauge theory dual to the Type IIB string theory in 
ten-dimensional pp-wave background can be thought to `live' on an 
{\it Euclidean} subspace spanning four of the eight transverse
coordinates. We then show that light-cone time evolution of the string is 
identifiable as the RG flow of the gauge theory --- a relation
facilitating `holography' of the pp-wave background. The `holography' 
reorganizes the dual gauge theory into theories defined over Hilbert 
subspaces of fixed R-charge. The reorganization breaks the 
SO(4,2)$\times$SO(6) symmetry to a maximal subgroup SO(4)$\times$
SO(4) spontaneously. We argue that the low-energy string modes may be
regarded as Goldstone modes resulting from such symmetry breaking pattern.
\vspace*{0.5cm}

\baselineskip=18pt
\newpage

\section{Overview}
It was known for some time that there is a certain limit, the 
so-called
Penrose limit, in which 
any spacetime which solves the Einstein's field equation
is reduced to a plane-wave background \cite{penrose}. Roughly speaking, the 
plane-wave background refers to spacetime close to a null geodesic. 
This assertion was extended to supergravity backgrounds \cite{guven}, 
involving, in addition to the metric, dilaton, p-form gauge fields, and 
fermionic superpartners. It was also realized \cite{bfhp} that 
maximally supersymmetric pp-wave backgrounds \cite{kowa,fp,bfp2,kg,meesen} 
are obtainable as the Penrose limits of the $AdS_p \times S^q$ backgrounds in 
ten-dimensional IIB supergravity and eleven-dimensional supergravity. 
Remarkably, the first-quantized superstring is exactly solvable 
in the pp-wave background \cite{metsaev, mettseyt}, as the
Green-Schwarz string action is quadratic in the worldsheet variables.

Recently, Berenstein, Maldacena and Nastase (BMN) \cite{bmn} argued that 
IIB string theory on such a pp-wave background with eight
transverse directions is dual to the large R-charge sector of ${\cal N} =4$ 
supersymmetric gauge theory in the large N limit. 
They identified a certain class of
long supermultiplet operators in the gauge theory with various string states.
By summing over a class of Feynman diagrams, they claimed that anomalous 
contributions to the scaling dimension of these operators indeed reproduce 
the dispersion relations predicted by the light-cone quantization. More 
significantly, they proposed a concrete construction of the ten-dimensional
string in terms of the four-dimensional gauge theory variables. 
If correct, the construction marks a 
significant progress beyond the AdS/CFT correspondence \cite{malda}, as 
it provides a dictionary for associating the gauge theory operators for not 
just supergravity modes, but for higher string modes as well. 
The BMN proposal has been also extended to backgrounds with less
supersymmetry \cite{mukhigomis, orbifolds, diff}.

In this paper, we substantiate aspects of the BMN proposal.
Specifically, we clarify holographic relations between the bulk
string states and the boundary gauge theory operators. In doing so, we
emphasize the crucial role played by the choice of the gauge theory
vacuum, on which both the superconformal symmetry and the R-symmetry
are spontaneously broken.  In section 2, we contrast the bulk-boundary
relations displayed in the AdS/CFT correspondence and those in the
pp-wave / gauge theory correspondence. In section 3, we illustrate this
by working out the profile of the supergravity modes in the pp-wave
background.  In section 4, we elaborate the pattern of the
aforementioned spontaneous breaking of conformal and R-symmetries.  We
emphasize that the dual gauge theory can be thought to be defined on
a \underline{Euclidean} four-dimensional space. We argue that 
holography relates the light-cone time in the pp-wave 
background to the renormalization group scale in the dual gauge 
theory. We show that this newly identified holography facilitates the 
nature of 
the string in terms of the dual gauge theory. In section 5, we 
discuss aspects of the enhanced supersymmetry in the dual gauge theory.
We conclude with remarks in section 6. 
\section{AdS/CFT versus PP-Wave/Yang-Mills}
In the AdS/CFT correspondence, the dual conformal field theory 
resides on the boundary of the AdS-space \cite{gkp,witten}, and the radial 
direction of the AdS-space plays the role of scale of the boundary theory 
\cite{reyyee,maldacena2,peetpolchinski,alvarezgomez,akhmedov,bvv}. 
Consider the global coordinates in $AdS_{d+1} \times S^{\bd + 1}$ 
space with metric 
\bea
ds^2 & = R^2 \left[ - \left( 1+\br^2 \right) d\bt^2 + 
\frac{d\br^2}{\left( 1+\br^2 \right) } + \br^2 d \Omega_{d-1}^2 
+ \left(1-\brh^2 \right)d\btheta^2 + \frac{d\brh^2}{\left( 1-\brh^2 
\right) } + \brh^2 d\Omega_{\bd - 1}^2 \right],
\label{eq:one}
\eea 
where the first and the second parts express the $AdS_{d+1}$ and
the $S^{\bd + 1}$ subspace, respectively. A bulk single particle state
of a given mass and spin, satisfying the classical field equation, is
specified by several ``momenta'': angular momentum quantum numbers
$\left(\ell,m_1,\cdots m_{d-2} \right)$ for the $S^{d-1}$ part in
$AdS_{d+1}$ space and $ \left(\bl,\bm_1 \cdots \bm_{\bd} \right)$ for
the $S^{\bd +1} $, respectively, and a principal quantum number $n$
for the remaining radial coordinate $\br$ in $AdS_{d+1}$ space.  The
bulk energy $\omega$ is then given in terms of these quantum numbers
by a dispersion relation. In the dual conformal field theory, we have
composite operators $\cO_{\{ \bl,\bm \} } \left(\bt,\phi_1 \cdots
  \phi_{d-1} \right)$, where $(\bt, \phi_1, \cdots, \phi_{d-1})$
denote coordinates of ${\bf R} \times S^{d -1}$. These operators are
decomposable into Fourier modes with a given energy $\omega$ and
$S^{d-1}$ spherical harmonics $\left(l,m_1, \cdots, m_{d-2} \right)$.
The remaining quantum numbers $\left(\bl, \bm_1, \cdots,
  \bm_{\bd}\right)$ are encoded in the structure of the operators. For
instance, in $AdS_5 \times S^5$, the $S^5$ quantum numbers are encoded
in the manner the six Higgs fields $\Phi^1, \cdots, \Phi^6$ of the
${\cal N}=4$ gauge theory appear in the operator. As a concrete
example, a bulk dilaton mode in $AdS_5 \times S^5$ with $S^5$ angular
momentum $\left(\bl,\bm_1, \cdots, \bm_{\bd} \right)$ is described by
a set of chiral primary operators whose bosonic component is given by
\be 
{\rm Tr} \, \left[F_{mn} F^{mn} \Phi^{(i_1} \cdots \Phi^{i_\bl )}
\right] (\bt,\phi_i, \cdots, \phi_{d-1}),
\label{eq:two}
\ee
in which the indices $i_1 \cdots i_\bl$ are decomposed into irreducible
representations of SO($\bd$). $F_{mn}$ denotes the gauge field strength.
Likewise, chiral primary operators
\be
{\rm Tr} \left[ \Phi^{(i_1} \cdots \Phi^{i_\bl )} \right] (\bt,\phi_i,
\cdots, \phi_{d-1})
\ee
describe modes of a linear combination of the four-form self-dual potential 
and the trace of the longitudinal graviton in ten-dimensions.

One can obtain the Penrose limit of Eq.(\ref{eq:one}) along a generic
null geodesic as follows. Boost along the two isometry directions:
\bea
t & = & ~\cosh \alpha ~\bt - \sinh \alpha ~\btheta \nn \\
\theta & = & - \sinh \alpha~\bt + \cosh \alpha~\btheta,
\label{eq:three}
\eea
and rescale two `radial' and light-cone coordinates:
\bea
r  =  R \br, \qquad \rho = R \brh \qquad {\rm and} \quad
x^\pm  =  {R \over {\sqrt 2}}(\theta \pm t).
\label{eq:four}
\eea
Then, take the limit
\bea
R \rightarrow \infty \qquad {\rm and} \qquad 
\alpha \rightarrow \infty, 
\label{eq:fivea}
\eea
while holding
\bea
x^\pm, \,\, r, \,\, \rho & = & {\rm fixed} \qquad
{\rm and} \qquad {e^\alpha \over {\sqrt 2} R} \equiv \mu = {\rm fixed}.
\label{eq:five}
\eea
The resulting spacetime is then reduced to
\bea
ds^2 & = & 2dx^+ dx^- -  \mu^2(r^2 + \rho^2) (dx^+)^2 + dr^2 + r^2 d\Omega_{d-1}^2
+ d\rho^2 + \rho^2 d\Omega_{\bd-1}^2 \nn \\
& = & 2dx^+ dx^- - \mu^2(\vx^2 + \vy^2) (dx^+)^2 + d\vx \cdot d\vx
+ d\vy \cdot d\vy,
\label{eq:six}
\eea
where we have defined transverse coordinates $\vx$, $\vy$ which describe the
${\bf R}^{d}$ made out of $r$ and $ S^{d-1}$, and ${\bf R}^{\bd}$ 
made out of $\rho$ and $S^{\bd-1}$, respectively. Even though the metric exhibits 
SO($d+\bd$) isometry, it turns out the RR five-form field
strengths break it to SO(d)$\times$SO($\bd$).

A novel feature of the pp-wave background is that the single particle bulk
states are now given in terms of certain harmonic oscillator quantum numbers
$(n_1 \cdots n_{d})$ and $(m_1 \cdots m_{\bd})$ for a given value of the
momentum conjugate to $x^-$, which we call $p_- \equiv 2 p^+$. The light-cone 
energy $p_+ \equiv 2 p^-$ is then given by a dispersion relation.
We will illustate this later in this section.
According to the BMN proposal, with these harmonic oscillator quantum 
numbers, the chiral primary operators dual to a single-particle bulk
state with the lowest light-cone energy, which turns out to be a 
linear combination of the self-dual RR four-form potential and trace
of the graviton, take the form of `$Z$-string':
\bea
\sum {\rm Tr} \left[ Z \cdots Z Z~(D_{i_1} Z)~ZZ\cdots ZZ~\Phi^{a_1}~ZZ
  \cdots ZZ~(D_{i_2} Z)~ZZ \cdots ZZ~ \Phi^{a_2}~ZZ \cdots \right].
\label{eq:seven}
\eea
Here,  along a string of $J$ factors of  $Z \equiv (\Phi^5 + i \Phi^6)$, 
one distributes $n_i$ insertions of $(D_i Z)$ and $m_a$ insertions of 
`transverse' Higgs fields $\Phi^a \, (a=7,\cdots 10)$. Then,
$\Phi^5$ and $\Phi^6$ are the two remaining, `longitudinal' Higgs
fields in the ${\cal N}=4$ gauge theory. 
The sum is over all distinct (up to cyclic permutation) locations of
the operators $D_i Z$ and $\Phi^a$ in the string of $Z$'s. 
The light cone momentum $p_- = - i {\partial \over \partial x^-}$ 
and the light cone energy $p_+ = i{\partial \over \partial x^+}$ are 
related to the dimension $\Delta$ and $J$ of the operator Eq.(\ref{eq:seven})
by the relations 
\bea
p_- & = & {1\over 2\mu R^2} (\Delta + J) = 
{1\over 2\mu R^2}\left( 2J + \sum_{i=1}^d n_i + 
\sum_{a=1}^\bd m_a \right) \nn \\
p_+ & = & \mu (\Delta - J) = \mu \left(\sum_{i=1}^d n_i + \sum_{a=1}^\bd m_a
\right).
\label{eq:pminus}
\eea
On the other hand, the coordinate transformations, 
Eqs.(\ref{eq:three}-\ref{eq:five}), imply
\bea
p_- & = & {1\over 2\mu R^2} \left( i{\partial \over \partial {\bt}}
- i{\partial \over \partial {\btheta}} \right) \nn \\
p_+ & = & \mu \left( i{\partial \over \partial {\bt}}
+ i{\partial \over \partial {\btheta}} \right).
\label{eq:pminustwo}
\eea
As $J = - i{\partial \over \partial {\btheta}}$, Eq.(\ref{eq:pminustwo}) is
consistent with Eq.(\ref{eq:pminus}) since
$\Delta = i{\partial \over \partial {\bt}}$ \footnote{Here we
assume that (\ref{eq:seven}) means that the operator is
evaluated at the origin of a ${\bf R}^4$}.

For other single-particle supergravity states such as the dilaton,
one needs to insert an operator $F_{mn}F^{mn}$
inside the $Z$-string. For higher string-mode states, 
each term in the sum of Eq.(\ref{eq:seven}) 
should be weighted by a phase-factor, which depends on
the locations of various operators in the $Z$-string. 

\underline{Note} that {\em all} the bulk quantum numbers appear in the 
structure of the dual gauge theory operators, Eq.(\ref{eq:seven}). This is
in sharp contrast to the AdS/CFT correspondence, where only \sl half \rm 
of the quantum numbers reside in the operator structure. There, the remaining
{\sl half} were encoded as dependence of the operator on coordinates of 
the four-dimensional spacetime, the boundary of $AdS_5$, on which the 
dual gauge theory resides.
Evidently, the operators in Eq.(\ref{eq:seven}) cannot be regarded as 
functions of the coordinates of the four-dimensional spacetime, as that
would result in more quantum numbers than needed for specifying a
given single particle supergravity/string state in the bulk.

In subsequent sections, we will argue that the gauge theory dual
to the pp-wave background Eq.(\ref{eq:seven}) can be thought of
as ``living'' in a  \underline{Euclidean}
four-dimensional space ${\bf R}^4$ spanned by the $\vx$ coordinates. 
The precise form of the dual gauge theory operators are
then given in terms of the Hermite transformation of local operators
defined on ${\bf R}^4$. The fact that
this space has to be Euclidean, rather than Minkowski spanned by
light-cone coordinates and part of ${\bf R}^4$, follows from the 
correspondence between the operators Eq.(\ref{eq:seven}) and the
one-particle states of the bulk supergravity/string theory. The 
latter states are described in terms of $(d + \bd)$ set of simple
harmonic oscillator operators with indices in a Euclidean space. 
For a string theory defined in the bulk, these oscillators also carry
a label for the level number \cite{metsaev}.
As we will see, this observation leads naturally to an interpretation of
$x^+$ as the holographic bulk coordinate in the Penrose limit, so that 
evolution in $x^+$ in the bulk generates scale transformation in the dual 
gauge theory. 

We will argue that in the gauge theory 
selecting a sector with fixed SO(2) charge $J$ is 
tantamount to a spontaneous
breaking of the conformal group SO(4,2) to SO(4) and the R-symmetry
group to SO(4) as well. The low-energy fluctuations are then
the Goldstone modes of the broken symmetries. The representation
of these operators in terms of Hermite transforms then follows
automatically.

\section{Supergravity Modes in pp-Wave Backgrounds}
Consider a minimally coupled, massless scalar field $D$ whose
field equation is given in the global coordinates Eq.(\ref{eq:six}) as
\be
\left[2 \partial_+ \partial_+
+ \mu^2 (\vx^2 + \vy^2) \partial^2_-
+ \sum_{i=1}^{d} \partial^2_{{\vx}^i}
+ \sum_{a=1}^{\bd} \partial^2_{{\vy}^a} \right]
\, D\left(x^+,x^-, \vx, \vy \right)
= 0.
\label{eq:aone}
\ee
The normal-modes with $p_- > 0$ are given by
\bea
D_{p_+,p_-, {\bf n}, {\bf m}} \left(x^+, x^-, \vx, \vy \right)
&=& e^{- {1\over 2} \mu p_- ({\bf x}^2 + {\bf y}^2} )
\prod_{i=1}^d H_{n_i} \left( \sqrt{\mu p_-} {\bf x}^i \right)
     \prod_{a=1}^{\bd} H_{m_a} \left( \sqrt{\mu p_-} {\bf y}^a \right)
\nn \\
&\times& \exp \left( ip_-x^- + p_+x^+ \right),
\label{eq:atwo}
\eea
where $H_{\bf n}({\bf x}), H_{\bf m} ({\bf y})$ denote the Hermite 
polynomials~\footnote{Note that, due to the
harmonic potential provided by the second term in Eq.(\ref{eq:aone}),
there is no real distinction between normalizable and non-normalizable
modes for $p_- > 0$. The modes with $p_- =0$ are not ${\cal L}_2$ 
normalizable, but are $\delta$-function normalizable. }.
The states of this scalar field theory 
are therefore created from the bulk Fock-space vacuum by creation 
operators ${\bf a}^\dagger ({\bf n}_i, {\bf m}_a, p^+)$ in the light-cone 
quantization. In a first-quantized theory of particles in this
background, states are created by creation operators 
${\bf c}^{i\dagger}, \,\, {\bf c}^{a\dagger}$:
\bea
{\bf c}^i = ({\bf p}_i + i {\bf x}^i )/\sqrt{2}  
\qquad &{\rm and}& \qquad 
{\bf c}^{i \dag} = ({\bf p}_i - i {\bf x}^i)/\sqrt{2},
\nn \\
{\bf c}^a = ({\bf p}_a + i {\bf y}^a )/\sqrt{2}  
\qquad &{\rm and}& \qquad 
{\bf c}^{a \dag} = ({\bf p}_a - i {\bf y}^a)/\sqrt{2},
\label{eq:cdef}
\eea
where the indices $i,a$ refer to the transverse directions along
${\bf R}^d \times {\bf R}^\bd$, spanning a $(d + \bd)$-dimensional
transverse space. 
The bulk dispersion relation is then given by
\be
2p^{-} = p_+ = \mu
\left(\sum_{i=1}^d n_i + \sum_{a=1}^\bd m_a + {1\over 2}(d +\bd) \right).
\label{eq:athree}
\ee
Note that the value of $p_+$ is independent of the value of $p_-$. This
is because the supergravity modes are massless.
For massive, string oscillation fields, the dispersion relation
depends explicitly on $p_-$. The sum over zero-point energies is
standard. 
We will see that, for the explicit example of ten-dimensional pp-wave 
background, this zero-point energy is precisely what is required for
precise correspondence with  appropriate operators in the dual gauge theory.
We note, for future reference, that the dispersion relation for the
mode which is a linear combination of the four form RR potential and 
the trace of the longitudinal graviton does not contain this zero-point 
fluctuation.

As we will elaborate more, the dual gauge theory has operators
which are Hermite transforms of local operators defined on the ${\bf R}^4$
spanned by ${\bf x}$.
\bea
{\cal O} [{\bf n}]
={\tt H.T.} \left[ {\cal O} \right],
\label{dilaton}
\eea
where the Hermite transform of a generic operator ${\cal O} (\vx)$ on 
${\bf R}^d$ is defined as
\bea
{\tt H.T.} \Big[ {\cal O} \Big] = {1 \over \cal N}
\int d \mu[\vx] \prod_{i=1}^d H_{n_i} \left(\sqrt{\mu p_-} x^i \right)
\,\, {\cal O}(\vx),
\nonumber
\eea
where ${\cal N}$ is a normalization factor, and the measure is given by
\bea
d \mu[\vx] := d^d \vx \, e^{ - {1 \over 2} \mu p_- {\bf x}^2}.
\nn
\eea
Using the standard recursion relation for Hermite polynomials, the Hermite
transform can be reduced to expressions involving derivatives acting
on ${\cal O}$ and no factors of the Hermite polynomials.

It may be useful to formally define operators on a 
${\bf R}^d \times {\bf R}^{\bd}$ by introducing a set of fiducial
coordinates ${\bf y}$ for the ${\bf R}^\bd$. Performing the Hermite transform 
on this eight-dimensional space,
\bea
{\tt H.T.} \Big[ {\cal O} \Big] = {1 \over \cal N}
\int d \mu[\vx] d \mu[\vy]  
\prod_{i=1}^d H_{n_i} \left(\sqrt{\mu p_-} x^i \right)
\prod_{a=1}^{\bd} H_{m_a} \left( \sqrt{\mu p_-} y^a \right) 
\,\, {\cal O}(\vx, \vy).
\nn
\eea
Using recursion relations obeyed by the Hermite polynomials, one can then 
express the Hermite transform in terms of derivatives
with respect to ${\bf y}$, which in turn become commutators with Higgs
fields inside the operator.

\subsection{Dilaton}
In the ten-dimensional pp-wave background, the dilaton field equation
take the same form as Eq.(\ref{eq:aone}). Thus, the light-cone energy 
spectrum of the dilaton state is given by
\bea
E_{\rm dilaton} = \mu
\left( \sum_{i=1}^4 n_i + \sum_{a=1}^4 m_a + {1 \over 2} (4
+ 4) \right).
\eea
According to the BMN proposal, the light-cone energy (measured in unit of 
$\mu$) ought to match with $(\Delta - J)$ of a gauge theory operator
dual to the dilaton.
A single insertion of $F_{mn} F^{mn}$, which carries $\Delta =4$
and $J = 0$, inside the $Z$-string in Eq.(\ref{dilaton}) is precisely
what we need to match the zero-point light-cone energy. Interestingly, in
providing the requisite zero-point energy $(4 + 4)/2 = 4$,
four-dimensionality of the internal space ${\bf R}^{\overline 4}$ has
played a crucial role. 

The single particle dilaton ground-state 
corresponds to 
${\bf n} = {\bf m} = 0$. For the states with higher energy, using the
recursion relation of Hermite polynomials, we deduce that the
corresponding operators are precisely insertions of the `transverse'
Higgs fields and covariant derivatives, viz. a set of operators of the
form
\bea
{\rm Tr} \left[F_{mn} F^{mn} ZZ \cdots ZZ \Phi^{a_1} ZZ \cdots ZZ
(D_i Z) ZZ \cdots ZZ \Phi^{a_2} ZZ \cdots \right].
\nn
\eea

\subsection{Longitudinal Graviton and Four-Form Potential}
The pp-wave background is supported by a homogeneous RR 5-form field strength
\bea
F_{+1234} = + F_{+5678} = \mu,
\nn
\eea
giving rise to Eq.(\ref{eq:one}) through the Einstein's field equation. 
As such, degrees of freedom of the graviton and the four-form RR potentials 
would mix each other. More precisely, expanding Type IIB supergravity
field equations of the metric and the RR four-form potential to linear
order fluctuations, $h_{\mu \nu}, c_{\mu \nu \alpha \beta}$, and taking
the light-cone gauge $h_{\mu - } = 0, c_{\mu \nu \alpha - }=0$, 
we find that the mixing takes
place between the traces of the graviton and the scalars of the RR
four-form potential. We thus denote these modes as
\bea
h := h_{ij} \delta^{ij} \qquad &{\rm and}& \qquad 
c := {1 \over 4!} \epsilon^{ijkl} c_{ijkl} \nn \\
\overline{h} := h_{mn} \delta^{mn} \qquad &{\rm and}& \qquad 
\overline{c} := {1 \over 4!} \epsilon^{mnpq} c_{mnpq}.
\nn
\eea
These fields are singlets of the two SO(4)'s 
on ${\bf R}^4 \times {\bf R}^{\overline 4}$, respectively.

Then, the linearized field equations exhibiting the mode mixing are given
by
\bea
\Delta_{\rm L} h - 16 \mu \partial_- c &=& 0 \nn \\
\nabla^2 c - 2 \mu \partial_- h &=& 0,
\nn
\eea
where $\Delta_{\rm L}$ stands for the Lichnerowitz operator for the 
spin-2 graviton. 
Utilizing the fact that $ \left( \Lambda_{\rm L} h \right)_{ij} = 
-{1 \over 2} \nabla^2 h_{ij}$ and diagonalizing the two coupled equations,
we obtain scalar-mode field equations
\bea
\left[ \nabla^2 - 8 i \mu \partial_- \right]
\left( h + 4 i c \right) = 0
\label{fourform}
\eea
and its complex conjugated equation for $(h - 4i c)$. 
Exactly the same set of equations hold for ${\bar h}$ and ${\bar c}$ as well.

The field equation Eq.(\ref{fourform}) is soluble exactly as in the
dilaton case. We find that the light-cone spectrum of the $(h + 4i c)$
`complex scalar' field is given by
\bea
E_{Z-{\rm scalar}} &=& \mu \left( \sum_{i=1}^4 n_i + \sum_{a=1}^4 m_a
+ {1 \over 2} (4 + 4) \right) - 4 \mu \nn \\
&=&  {\mu} \left( \sum_{i=1}^4 n_i + \sum_{a=1}^4 m_a \right) \ge 0.
\nn
\eea
On the right-hand side of the first expression, the first and the second 
terms are contributions from $\nabla^2$ and $- 8 i \mu \partial_-$, 
respectively. Evidently, the
zero-point energy arising from fluctuations along the eight transverse 
directions is cancelled precisely by the classical contribution $- 4 \mu$ 
to the light-cone energy. Hence, along with the second set of complex
`scalar' field $({\bar h} + 4 i {\bar c})$, we conclude that there are two bulk
`scalar' modes yielding the minimum of the light-cone energy to be zero. These bulk `scalar' fields are then identified
with the dual gauge theory operators
\bea
{\rm Tr} \left[ ZZ \cdots ZZ \cdots ZZ \right],
\nn
\eea
viz. the Z-string, first introduced by BMN.

In contrast, the complex-conjugate `scalar' fields $(h - 4i c)$
and $({\bar h} - 4 i {\bar c})$ 
are subject to the classical contribution $+2 \mu$ to the light-cone energy.
It implies that the minimum of the light-cone energy is
$+8 \mu$, instead of 0, 
rendering the corresponding dual gauge theory operator involving eight
powers of $\Phi^a$'s distributed along the $Z$-string. 

\section{Penrose Contraction, Spontaneous Symmetry Breaking \& Euclidean
Dual Gauge Theory}
We now turn to the dual ${\cal N}=4$ supersymmetric gauge theory. 
This theory is invariant under SO(4,2) $\otimes$ SO(6), where SO(6) refers 
to the internal R-symmetry. Denote the generators of 
SO(4,2) as $J_{AB}$ with $A,B = 1, \cdots, 6$, where $5,6$ are the 
directions with negative signature,  and those of 
SO(6) as $J_{UV}$ with $U,V = 7, \cdots, 12$.
In terms of $J_{AB}$, the generators of the conformal group are
\be
J_{ij}, \qquad P_{i} = J_{5i} + J_{6i},  \qquad K_{i} = J_{5i} - J_{6i},
\qquad D_{1} = J_{56}
\ee
with $ i,j = 1, \cdots, 4$. 
The same can be done for the generators of SO(6), and we
define $J_{ab},\, P_{a}, \,K_{a}, \,D_{2}$ accordingly, 
where $a,b =7, \cdots, 10$ and $D_{2}=
J_{11,12}$. 

Let us now assume that there exists a vacuum state, on which
the SO(4,2)$\otimes$ SO(6) is broken spontaneously to SO(3,1)
$\otimes$ SO(4), viz. standard symmetry breaking pattern 
preserving Lorentz plus `transverse'
internal symmetries. The number of generators of broken symmetries
is eighteen, viz. nine nonlinearly realized symmetries for each
product group. The generators of the broken
symmetries are
$P_{i}, \, K_{i}, \, P_{a}, \, K_{a}, \, D_{1}, \,D_{2}$ and the
generators of the unbroken symmetries are
the $J_{ij}$ for the Lorentz group SO(3,1) and the $J_{ab}$ for the
internal symmetry group SO(4). One easily finds that generators of
the  broken symmetries satisfy the following commutation relations
\bea
[P_{i},K_{i}] = D_{1} \qquad {\rm and} \qquad [P_{a},K_{a}] = D_{2}.
\nn
\eea
These commutation relations are very suggestive. If one were to put aside the
fact that $D_{1}$ and $D_{2}$ do not commute with the $P$'s and $K$'s,
one may try to interpret the previous commutation relations as defining
two Heisenberg algebras {\tt h}(4)$\oplus$ {\tt h}(4), each 
one with eight generators, for which the $D_{1}$ and $D_{2}$ as the two 
central extensions. This interpretation, as it stands, is not viable if 
one just considers the standard symmetry breaking pattern 
SO(4,2) $\otimes$ SO(6) to SO(3,1) $\otimes$ SO(4) : $D_{1}$ and 
$D_{2}$ are not central terms and we can not organize the generators of
the broken symmetries in terms of two Heisenberg algebras. 
It is precisely at this point where the existence of
supergravity/string  duals and concept of the Penrose contraction can
help us to define a different pattern of the symmetry breaking. 

\subsection{Penrose Contraction}
As is well known, the symmetry algebra SO(4,2) $\otimes$ SO(6) of
${\cal N}=4$ gauge theory are realizable as isometries of the
$AdS_{5} \times S^{5}$ spacetime. The Penrose limit recapitulated in
secton 2 preserves the total number of Killing vectors but can change
their algebraic relations. In particular, if we perform the Penrose
limit on a generic light geodesic in  $AdS_{5}\times S^{5}$ the
Killing vectors define the algebra
[{\tt h}(4)$\oplus${\tt h}(4)]$\oplus${\tt so}(4)$\oplus${\tt so}(4), 
where the bracket is to emphasize the fact that two Heisenberg 
algebras share the same central extension. The extra Killing
vector defines an outer-automorphism of the Heisenberg algebras. We
interpret
the Penrose limit as defining a sort of spontaneous symmetry breaking from
SO(4,2)$\otimes$SO(6) to SO(4)$\otimes$SO(4) with the eighteen generators of
the broken
symmetries defining the two Heisenberg algebras {\tt h}(4)'s and the 
outer-automorphism. 

As the simplest illustration, consider $AdS_{2}\times S^{2}$, relevant
for the near-horizon geometry of four-dimensional BPS black holes. In
this case, the symmetry group is SO(1,2)$\otimes$ SO(3) with 
six generators that we will denote $P_{1},\, K_{1},\, P_{2},\, K_{2},\, D_{1}, \, D_{2}$.
They satisfy, in particular, $[P_{1},K_{1}] = D_{1}$ and $[P_{2},K_{2}]= D_{2}$
\footnote{ In
\cite{bfp2} the generators $P_{i}\, K_{i}, \,D_{i}$ correspond respectively to
the
Killing vectors $E_{i},E^{*}_{i}, \epsilon_{i}$.}. In the Penrose limit,
$P_{i},K_{i}$
become the generators of two Heisenberg algebras and the $D_{i}$'s produce
the
common central term and the outer automorphism. In fact, denoting the
Penrose scaling by $\Omega$, we get 
$D_{i}(\Omega) = d^{i,0} \Omega^{-2} + d^{i,1} + d^{i,2} \Omega^{2} +... $
with
$d^{1,0} = d^{2,0}$. The central term is defined by $d^{i,0}$ and the
outer-automorphism by $\left(d^{i,1} - d^{i,2} \right)$. 
Expansion of
$D^{i}(\Omega)$ is then interpretable 
as a perturbative expansion in powers of the Penrose scaling 
parameter, $\Omega$.

An important aspect of the Penrose limit in the case of $AdS_{5}
\times S^{5}$ considered by BMN is that
the unbroken symmetry is SO(4)$ \otimes $SO(4). In other words, if we want
to use the
Penrose contraction as a pattern of the symmetry breaking for the dual
${\cal N}=4$ gauge theory,
we should assume that the vacuum is invariant not under the Lorentz group
but under a rotation group in a four-dimensional Euclidean space. 
Insight to this possiblity can be gained by recalling aspects of
spontaneous conformal symmetry breaking, studied thoroughly some time
ago \cite{salam}. The idea was to assume an underlying theory
invariant under the conformal group and, after spontaneous conformal
symmetry breaking, to study the
low-energy physics of the corresponding Goldstone bosons. The first
peculiar
aspect of the spontaneous conformal symmetry breaking,  SO(4,2) to
SO(1,3), 
is that the generators of translations are part of the broken symmetries. 
Being so, only the generators of special conformal
transformations
and dilatations were considered \cite{salam} 
as real Goldstone bosons. A consequence 
of this is that these Goldstone bosons, contrary to the standard case, are
not
massless as the broken symmetries do not commute with the Hamiltonian,
viz. with
translations in time. In the Penrose contraction, we are facing a similar
problem. If we consider ${\cal N}=4$ gauge theory and the
standard spontaneous 
breaking pattern to SO(3,1) $\otimes$ SO(4), we are including among the broken
symmetry generators, 
the translation generators in physical time as well as the spatial
translation generators. If
we try to understand this breaking in the old-fashioned approach
\cite{salam}, we need to
organize the eighteen broken symmetries into a set of
nine massless Goldstone bosons, corresponding to the spontaneous 
breakdown of the internal symmetry SO(6), five massive Goldstone bosons
corresponding
to the special conformal transformations and dilatations that
do not commute with the Hamiltonian and four translations. This is
certainly not
the picture we get if we use Penrose contraction. In the Penrose
contraction,
we organize the eighteen broken symmetries into a Heisenberg algebra 
{tt h}(8) and an outer-automorphism. What now remains is a  concrete 
interpretation of the Heisenberg algebra {\tt h}(8) and the 
outer-automorphism entirely within the dual gauge theory formulation. 

\subsection{Dual Gauge Theory is Euclidean}
Let $\Phi_{i}$
$i = 5...10$ be the Higgs fields of ${\cal N}=4$ super Yang Mills theory. 
Following BMN, define the field $Z= (\Phi_{5}+ i \Phi_{6})$, and denote
by  $J$ the SO(2) R-charge corresponding to rotations in the internal
(5,6)-plane. Consider decomposing the gauge theory 
Hilbert space into infinite
towers of Hilbert subspaces of definite $J$ quantum number. Evidently,
on each subspace, Fock-space ``ground state'' breaks the internal
SO(6) spontaneously to SO(4). We denote the Fock-space vacuum with 
R-charge equal to $J$ as $\vert 0 \big>_J$. 
We will be interested in the Hilbert space of quantum fluctuations
around this vacua. The first thing to be done
is characterizing the state $\vert0\big>_J$. The simplest way to define this
state is, in radial quantization,
\bea
{\rm Tr} \left(Z^{J} \right)(\vx=0) \vert 0 \big>_{\rm YM},
\nn
\eea
where $\vert 0 \big>_{\rm YM}$ refers to the perturbative vacuum of the
dual ${\cal N}=4$
gauge theory. The operators should be considered as functions of
the coordinates of a four dimensional  
\underline{Euclidean} space, ${\bf R}^{4}$, and is {\sl not} 
related a priori to
Euclideanized ${\cal N}=4$ super Yang-Mills  theory defined on ${\bf R}^3
\times {\bf R}_\bt$ after the Wick rotation. On ${\bf R}^4$, a \sl local 
\rm operator Tr$\left(Z^{J} \right)(\vx)$ is expandable in a complete basis 
of the Hermite polynomials
\be
{\rm Tr} \left(Z^{J} \right)(\vx) = \sum_{\{ {\bf n} \}} 
c_{\bf n} \prod_{i=1}^4
\left(e^{-\Lambda^2 x_i^2} ~ H_{n_i} (\Lambda x_i) \right),
\label{eq:defoper}
\ee
where $\Lambda$ is a scale defined within the dual gauge theory, 
which will be determined later. Thus, we can write
\be
\vert 0 \big>_J = c_{{\bf n} =0} \, \vert 0 \big>_{\rm YM}.
\ee
The Hilbert space of quantum fluctuations is generated by states
$\vert {\bf n} \big>_J = c_{\bf n} \vert 0 \big>_{\rm YM}$. 
For instance, we get:
\bea
\sum_{\ell} {\rm Tr}\left(Z^{\ell}(D_{i}Z)Z^{J- \ell}\right)(0)
\vert 0 \big>_{\rm YM} = \vert {\bf n}_{i}=1\big>_J.
\nn
\eea
One can define creation and annhilation operators $b_{0}^{i}$ and
$b_{0}^{i \dag}$ obeying the canonical commutation relation
$\left[ b_0^i , b_0^{j \dag} \right] = \delta^{i,j}$ such that
\bea
b_{0}^{i \dag } \vert 0 \big>_J  = \vert {\bf n}_{i}=1 \big>_J.
\eea
These operators generate the Heisenberg algebra {\tt h}(4).
As we are working in ${\cal N}=4$ gauge theory, we can 
also consider fluctuations with respect to the internal directions, namely,
\bea
\sum_{\ell}{\rm Tr} \left(Z^{\ell}\Phi_{a}ZZ^{J-\ell} \right)(0)
\vert 0 \big>_{\rm YM}
\eea
with $a = 7, \cdots, 10$. In the large-$J$ limit, one can represent these states
in terms of the same type of creation and annhilation operators as before,
viz. 
$\vert {\bf n}_{a}=1 \big>_J = b_{0}^{a \dag} \vert 0 \big>_J$. 
Both $b_{0}^{i}$ with $i= 1, \cdots, 4$ and $b_{0}^{a}$ with $a=7, \cdots, 10$
transform as vectors under the two SO(4)'s respectively. 
>From now on, we will denote them collectively as $b_{0}^{i},b_{0}^{i \dag}$
with $i=1, \cdots,4,7, \cdots, 10$. 
These operators generate the Heisenberg algebra 
{\tt h}(8). Note that this is true only in
the large-$J$ limit and for Euclidean gauge theory.
In view of the BMN proposal, 
it is quite natural to identify this Heisenberg algebra {\tt h}(8) with 
the similar Heisenberg algebra encountered in the Penrose contraction 
of SO(4,2)$\otimes$SO(6). 

The next step would be to identify, within the dual gauge theory, 
the physical meaning of both the central extension
and the outer-automorphism. In the original theory invariant under
SO(4,2)$\otimes$ SO(6), there are two generators of the symmetry algebra
that are of special importance, viz. the generator of dilatations of
the space-time coordinates and the generator $J$ of the SO(2) R-symmetry. 
Inferring from the discussion on the Penrose contraction in section 2, we 
should expect that both the central extension and the outer-automorphism are
associated with these two generators. In the dual gauge theory,
these generators have a clear physical meaning: 
generator of the space-time dilatations
defines {\sl scaling dimension} $\Delta$ of the operators, and the
generator $J$ defines the 
corresponding R-charge. Note that in the $AdS_5$ realization of $SO(4,2)$
the embedding coordinates $X^A (A = 1,\cdots 6)$ 
with $(X^1)^2 + (X^2)^2 - (X^3)^2 - \cdots (X^6)^2 = R^2$
are given in terms of the global coordinates $(\bt,\br,\phi_i)$ as
\bea
X^1 &=& R {\sqrt{1+\br^2}}~\cos~\bt \nn \\
X^2 &=& R {\sqrt{1+\br^2}}~\sin~\bt \nn \\
X^\alpha &=& R\br \omega^\alpha. \qquad \qquad (\alpha = 3,\cdots, 6)
\label{eq:hone}
\eea
Here, $\omega^\alpha$ denote the embedding coordinates of a unit $S^3$.
The standard "dilatation" generator of the SO(4,2) group, which is $J_{12}$,
generates translation in global time $\bt$. In the Penrose limit,
$\br \rightarrow 0$, while $r = R \br$ held fixed. Thus, $X^\alpha \,\, 
(\alpha = 3, \cdots, 6)$ are ${\cal O}(1)$ and become {\it unconstrained},
while $X^1,X^2$ are of ${\cal O}(R)$. The dual gauge theory is now defined
on the Euclidean plane $X^3, \cdots, X^6$ and $J_{12}$ generates the 
scale transformations on this ${\bf R}^4$-subspace as in the standard 
realization of the SO(4,2) group. 
Let us denote the eigenvalues of these generators,
for a given operator, as $\Delta$ and $J$, respectively.
For the state $\vert 0 \big>_J $, we have
\bea
\Delta \vert 0 \big>_J = J \vert 0 \big>_J \qquad
{\rm and} \qquad  J \vert 0 \big>_J = J \vert 0 \big>_J,
\nn
\eea
while, for states of type $b_{0}^{i \dag}...b_{0}^{i \dag}
\vert 0 \big>_J$, we have
$\Delta = (J+n)$, where $n$ refers to the number of $b^\dag$-oscillators
and $J=J$. 
Thus, on these
states, we have $(\Delta +J) = 2J +n$ and $(\Delta -J) = n$.
If we work in the large-$J$, large-N, and small-$n$
limits with
\bea
{4 \pi g^2_{\rm YM} N} := R^2 \rightarrow \infty, \qquad 
J^2 \rightarrow \infty \qquad {\rm and} \qquad 
\frac{R^2}{J^{2}} := g^2_{\rm eff} \rightarrow {\rm finite},
\label{limits}
\eea
we observe that, in appropriate units, 
${(\Delta +J)}\slash{R^2}$ becomes the true 
central extension commuting with the $b_{0}^{i}$ and  $b_{0}^{i
\dag}$ operators
\footnote{
Note that once we identify this term with the central extension of the
Heisenberg algebra, we need to normalize the  $b_{0}^{i}$ and  $b_{0}^{i
\dag}$ so that they obey the canonical commutation relations.},
 and that  $(\Delta-J)$ is simply the number operator
for the $b_{0}^{i}, \, b_0^{i \dag}$ oscillators and therefore is a true outer-automorphism
of the Heisenberg algebra.

In summary, built only upon Euclidean gauge theory residing on 
${\bf R}^{4}$ subspace, we succeeded in finding a
SO(4)$\otimes$SO(4) invariant vacuum and a representation of the
Heisenberg group H(8) in terms of creation and annhilation
operators acting on the Hilbert space of small quantum 
fluctuations. The corresponding outer-automorphism is just
the number operator. Note that the vacuum state $\vert 0 \big>_J$ is not
only invariant under SO(4)$\otimes$ SO(4) but also with respect
to the one-parameter group generated by the outer-automorphism. 

So far, we have considered only the modes which are chiral primaries.
The scaling dimension $\Delta$ of the corresponding operator is
\be
\Delta = \left( J + \sum_{i=1}^4 n_i + \sum_{a=1}^4 m_a \right),
\label{eq:dimone}
\ee
where there are $n_i$ insertions of $D_i Z$, and $m_j$ insertions of
$\Phi^j$.
Supersymmetry descendants of these would contain
factors involving the gauge field, as discussed in section 3. 
Consider, for example, the dilaton.
The operator dual to this should be the integral of
\bea
{\rm Tr} [F_{mn}F^{mn} \Phi^{(i_1}\cdots \Phi^{i_\bl )}] (t,\phi_i
\cdots \phi_{d-1}).
\nn
\eea
For such operators, the scaling dimension $\Delta$ is given by
\be
\Delta = J + \sum_{i=1}^4 n_i + \sum_{m=1}^4 m_a + 4.
\label{eq:dimtwo}
\ee
These relations are consistent with our interpretation
of the holographic coordinate.

The scaling dimension $\Delta$ of the dual operator is, however,
the eigenvalue of the operator $i \partial_\bt $ in the bulk theory. 
The R-charge is of course the eigenvalue of $ -i \partial_\btheta$.
These relations are in accord with the solutions of the bulk wave equations. 
Take the dilaton as an example. From Eq.(\ref{eq:athree}) with 
$d=4$ and $\bd = 4$, we have
\bea
2p^- = \mu \left( \Delta - J \right) = {\mu} \left(\sum_{i=1}^4 n_i 
+ \sum_{a=1}^4  m_a + 4 \right)
\label{eq:bone}
\eea
and find precise agreement with Eq.(\ref{eq:dimtwo}). From the bulk point of
view, the additive factor 4 appears as a zero point energy. From the
gauge theory viewpoint, this reflects the presence of $F_{mn}
F^{mn}$ in the operator. For the bulk mode which is a fluctuation
of the four-form RR potential, this zero point energy is absent, which
is consistent with the absence of any factor of the gauge field strength
in the dual operator. 
In the $J \rightarrow \infty $ limit, $\Delta \rightarrow \infty$ as well. 
However, $p_+ = \mu (\Delta - J)$ remains finite. 
This is the reason why, 
though it appears natural to consider $t$ as a
holographic direction from the bulk point of view, it is actually
more natural to consider $x^+$ as the
holographic direction from the gauge theory point of view.

Finally, it should be 
clear from the definition of the operators that one needs
to introduce a cutoff mass scale $\Lambda$. Inferring from the expressions for 
the normal modes of the bulk fields, it is natural to choose the scale 
to be given, up to ${\cal O}(1)$ numerical factor,  by
\be
\Lambda = \mu p_- \, .
\ee
At first sight, this identification appears strange. 
It would mean that one needs {\sl a priori} a different scale for each 
operator as $p_-$ is defined by $(\Delta + J)/ 2 \mu R^2$. However,
in the approximation adopted in Eq.(\ref{limits}), $n_i, m_a \ll J$. As
such,  $p_- \sim J / R^2 \sim {\cal O}(1)$. This implies that all the 
operators involved are governed universally by a common renormalization scale. 

\subsection{Where does the dual gauge theory ``live'' ?}

For the ten-dimensional pp-wave, the states created by the $c_{\bf n}$
defined in Eq.(\ref{eq:defoper}) are to be identified with the states
in the light-cone gauge quantized Type IIB string in the bulk created
by {\em transverse} oscillators. It is thus natural to identify one of
the transverse ${\bf R}^4$'s of the pp-wave with the space on which the
Euclidean gauge theory ``resides''. Furthermore, as the gauge theory
operator ${\rm Tr}(Z^J)$ does not contain any of the Higgs fields
$\Phi^7 \cdots \Phi^{10}$, the transverse ${\bf R}^4$ in question
would be identifiable with the ${\bf R}^4$ formed by the $X^\alpha,
\alpha = 3 \cdots 6$ in Eq.(\ref{eq:hone}).

To understand the meaning of this interpretation, let us trace back
the process of taking the Penrose limit.  The starting bulk theory,
the $AdS_5 \times S^5$ supergravity theory, is dual to ${\cal N}=4$
supersymmetric gauge theory, which is defined on Lorentzian manifold,
$S^3 \times {\bf R}_\bt$, where the $S^3$ and ${\bf R}_\bt$ denotes a
foliation leaf at $\br = \br_0$, with large $\br_0$, and the global
time $\bt$ in $AdS_5$.  One possible way to obtain a gauge theory on a
${\bf R}^4$ is by Wick-rotating the global time $\bt \rightarrow i
\tau$, and then performing the standard conformal transformation of
the resulting Euclidean $S^3 \times {\bf R}_\tau$.  We will refer the
euclidean space obtained in this way by ${\bf \hR}^4$ and its
coordinates by ${\bf z}^i$. One may then try to interpret the
derivatives $D_i$ appearing in the operator in Eq.(\ref{eq:seven}) as
derivatives with respect to the ${\bf z}^i$.  However, this cannot be
right if the insertion of a $D_i Z$ corresponds in the bulk to a state
obtained (in a first quantized description of particles) by applying
an operator ${\bf c}^{i\dagger}$ defined in Eq.(\ref{eq:cdef}). First,
the description of string or supergravity states in the light-cone
gauge does not, of course, involve any analytic continuation. We
should not perform any analytic continuation in the gauge theory as
well. Secondly, the index $i$ in ${\bf c}^{i\dagger}$ clearly refers
to a direction transverse to the pp-wave : in this case the transverse
${\bf R}^4$ is formed by $(\br, \phi_1, \cdots, \phi_3)$.  In other words,
the symmetry of the set of states created by ${\bf c}^{i\dagger}$
includes a SO(4) and this SO(4) does not involve the global time,
Wick rotated or not.

In concrete terms, our statement that the dual theory can be thought
of as living on the transverse ${\bf R}^4 = (\br,\phi_1, \cdots,
\phi_3)$ means that the derivatives which appear in
Eq.({\ref{eq:seven}) should be derivatives with respect to this
transverse ${\bf R}^4$. This provides a manifest linear realization
of the symmetries SO(4)$\times$SO(4) and the associated Heisenberg
algebra ${\tt h}(8)$.

However, the Penrose limit is supposed to be a sector of the original
Yang-Mills theory defined on the $S^3 \times {\bf R} = (\bt, \phi_1
\cdots \phi_3)$. How can we then regard the operators of this theory
to depend on $(\br, \phi_1, \cdots, \phi_3)$ ? The answer is provided by
the IR-UV connection in $AdS$ holography. In the gauge theory
composite operators are functions on the space on which the theory is
defined, $\cO ({\bf z}) = \cO( \bt,\phi_1, \cdots, \phi_3)$.  At the
quantum level, however, these local operators also depend
on the (position space) ultraviolet cutoff $a$ , or equivalently a
renormalization scale. So, these local operators should really be
expressed as $\cO ({\bf z};a) = \cO( \bt,\phi_1 \cdots \phi_3; a)$.
The AdS/CFT correspondence identifies this operator with a {\em bulk
operator} (e.g. for one of the supergravity fields) $\Phi
(\bt,\phi_1, \cdots, \phi_3, \br)$ evaluated at large $\br = \br_0 =
1 \slash a $. We can thus think of the transverse ${\bf R}^4 
= \{ {\vec {\bf x}} \}$ above as being formed by combining the
spatial coordinates $\phi_i$ with the scale $a$. 
In terms of the original Yang-Mills theory living on $S^3 \times 
{\bf R}_{\bt}$ derivatives
which appear in Eq.({\ref{eq:seven}) therefore involve
derivatives with respect to the scale.

Before considering any Penrose limit, it is natural to arrange `twist'
expansion of operators as in $\cO ({\bf z} = 0;a)$ and $D_{\bf z} 
\cdots D+{\bf z} \cO
({\bf z}=0; a)$, while, after the Penrose limit, the more convenient
`twist' expansion is with respect to ${\bf x}^i$, viz. $D_{\bf x} \cdots
D_{\bf x} \cO
({\bf x}=0, \bt)$.  Upon swapping $\br$ with the light-cone time, we
are also led to do so for the `holography' direction.  In this sense,
the light-cone time plays the role of the `holography' direction,
while $\br$ is now part of the Euclidean base space ${\bf R}^4$.

Evidently, the above two choices of `twist' expansions are intimately
related.  Schematically, the renormalization group equation of the
operator $\cO$ takes the form 
\bea 
a {d \over da} \cO(a) = \gamma \cdot \cO (a).  \nonumber 
\eea 
By virtue of the AdS/CFT correspondence the renormalization group 
equations of the gauge theory are related to the 
(suitably formulated) bulk equations of motion.
Replacing $1/a$ with the
radial coordinates $\br$, the renormalization group can then be
recasted as a first-order light-cone time evolution obeyed by a data
specified on ${\bf R}^4$ formed out of $\br$ and $(\phi_1, \cdots,
\phi_3)$. An implicit assumption in the two alternative `twist'
expansions is that the locality, which was automatically built in the
Minkowski gauge theory on $S^3 \times {\bf R}_\bt$, is maintained in
the Euclidean gauge theory on ${\bf R}^4$. This is highly nontrivial,
as, in the standard $AdS_5$ holography, map between the gauge theory
scale and the $AdS_5$ radial position is given in a complicated and
nonlocal manner. Apparently, in the Penrose limit, the map becomes
simplified dramatically and becomes local, as inferred from the precise
matching between the light-cone string dynamics and the Euclidean gauge
theory dynamics. 

Summarizing, we emphasize that the question regarding where the gauge 
theory dual to a pp-wave light-cone string lives is intimately
connected with the way we identify the string transverse oscillators
into the Hilbert space of quantum fluctuations of a fixed R-charge
ground-state in the gauge theory. Because we are focusing on each
Hilbert subspace of fixed R-chrage, the question is also intimately
tied with the realization of the pp-wave isometries in the gauge theory. 
The normal subgroup of the latter is SO(4)$\otimes$SO(4), where one
SO(4) originating from the kinematical conformal group in four
dimensions and the other from the internal R-symmetry. These
symmetries are manifest once operators and states in the dual 
gauge theory are formulated on the transverse ${\bf R}^4$. In doing
so, we interchange the role between the light-cone time $\bt$ and the
bulk radial coordinate $\br$, and now view the former as the
`holography' coordinates of the dual, Euclidean gauge theory. 
 
\subsection{Light-Cone Holography}
Once we have defined the outer-automorphism, call it $H$, we 
can trivially use it to define a one
parameter family of operators. In fact, we can introduce a formal parameter
$x^{+}$ as a conjugate variable to $H$ and define
\bea
\sum_{\ell} {\rm Tr}\left(Z^{\ell}(D_{i}Z)Z^{J-\ell}
\right)(0,x^{+}) \vert 0 \big> = e^{-i x^{+} H} 
b_{0}^{i \dag} \vert 0 \big>_J
. \nn
\eea
This is the prescription for introducing an extra 
\sl holographic \rm 
light-cone time coordinate in the 
Euclidean gauge theory. In fact, what we get is simply
\be
\sum_{\ell} {\rm Tr} \left(Z^{\ell}(D_{i}Z) Z^{J-\ell} \right)(0,x^{+})
\, \vert 0 \big>_{\rm YM} = e^{-i x^{+}(\Delta -J)}
\, \sum_{\ell} {\rm Tr} \left(Z^{\ell}(D_{i}Z) Z^{J-\ell} \right)(0)
\vert 0 \big>_{\rm YM}
\ee
and the outer-automorphism $H = i \partial_{x^{+}}$. If we wish, we can
also introduce an another holographic coordinate, say, $x^{-}$  conjugate 
to the central extension $C = (\Delta +J) /\lambda$ by
\bea
\sum_{\ell} {\rm Tr} \left(Z^{\ell} (D_{i}Z) Z^{J-\ell} \right)
(0,x^{+}, x^{-}) \vert 0 \big>_{\rm YM} =
e^{i x^{-}C} e^{- i x^{+} H} \, b_{0}^{i
\dag} \vert 0 \big>_J. 
\nn
\eea
This dependence on $x^{-}$ is trivial, as it is the same for all operators.
It depends only on the particular finite value we choose for 
$R^2/{J^{2}}$ in the double large-$J$ and large-$N$ limit. Once
we introduce the coordinate $x^{-}$, the central extension $C$ can be
represented as $-i \partial_{x^{-}}$.

We can readily make a contact with the spacetime Killing vectors in the
Penrose limit. To illustrate this, consider the case of ${\rm AdS}_{3}
\times S^{3}$. In the Penrose limit, the isometry group 
SO(2,2)$\otimes$SO(4) is contracted to [H(2)$\otimes$H(2)]$\otimes$SO(2)$
\otimes$SO(2). We are interested in the generators of the Heisenberg
groups H(2)'s:
\bea
P_{i}\left(x^{+}\right) &=&  \cos\left(\mu x^{+}\right) 
\frac{\partial}{\partial x^{i}} + \mu x^{i} \sin \left( \mu x^{+} \right)
\frac{\partial}{\partial x^{-}}, \nn \\
K_{i}\left(x^{+}\right) &=&  \sin\left(\mu x^{+}\right) 
\frac{\partial}{\partial x^{i}} - \mu x^{i} \cos\left(\mu x^{+}\right)
\frac{\partial}{\partial x^{-}}.
\nn
\eea
These are the generators that we should put in correspondence with
the operators $b_{0}^{i}$ and $b_{0}^{i \dag}$ in the dual gauge theory.
As discussed above, the operator 
$-i \partial_{x^{-}}$ plays the role of the
central extension $C$ of the Heisenberg algebra. In the standard 
coordinate-momentum notation, we can write
\bea
P_{i}\left(x^{+}\right) &=& p_i \, \cos\left(\mu x^{+}\right) +  
\mu C x^i \, \sin\left(\mu x^{+}\right)
\nn \\
K_{i}\left(x^{+}\right) &=& p_i \, \sin\left(\mu x^{+}\right) -  \mu C x^i \, 
\cos \left(\mu x^{+} \right),
\nn
\eea
where the parameter $C$ represents the central extension term, and
\bea
\left[p_{i},x^{j} \right]= - i \delta_{i}^{j}.
\nn
\eea
Evidently, we can write
\be
P_{i}(x^{+}) = e^{- i x^{+}H}\, P_{i}(0) \, e^{+ i x^{+}H}
\ee
for $H$ the light-cone Hamiltonian defined by $H = i \partial_{x^{+}}$. This
shows that the Hamiltonian $H$ is an outer-automorphism of
the Heisenberg algebra.

We close this section with comments. Quantum mechanically, 
the value $\Delta$ of the scaling dimension
is corrected by the contribution of anomalous dimensions, whereas 
the value $J$ of the R-symmetry charge remains unchanged.  As such, one
ought to expect quantum corrections to the outer-automorphism $(\Delta -J)$.
How is it corrected?
The interaction part of ${\cal N}=4$ gauge theory contains a term
of the type $ \frac{1}{2 \pi g^2_{\rm YM}} {\rm Tr} [Z,\Phi_a][Z,\Phi_a]$.
This term induces a non-vanishing transitions of the type:
\bea
\left< {\rm Tr} \left(Z^{\ell}\Phi_a Z^{J-\ell} \right)(x)\, {\rm Tr} \left(
Z^{\ell+1}\Phi_b Z^{J-(\ell+1)} \right)(0) \right> = 
\delta_{ab} \, 2\pi g^2_{\rm YM} N {\cal I}(x)\frac{1}{ 4 \pi^2 x^{2J-2}}
\eea 
and
\bea
\left<{\rm Tr} \left(Z^{\ell}\Phi_a Z^{J-\ell} \right)(x) \, {\rm Tr} \left(Z^{\ell}\Phi_b Z^{J-\ell)} \right)(0) \right> = \delta_{ab} \, 2 \pi 
g^2_{\rm YM} N \, {\cal I}(x) \frac{1}{4 \pi^2 x^{2J-2}}.
\nn
\eea
The function ${\cal I}(x)$ is given by
\bea
{\cal I}(x) = {1 \over 4 \pi^2} \log |x| \Lambda + {\rm finite},
\nn
\eea
where $\Lambda$ defines the ultraviolet cutoff.
Both contributions cancel each other, meaning that, for operators
of the type $\sum_{\ell} {\rm Tr} \left(Z^{\ell}(D_{i}Z) Z^{J-\ell} \right)$
or $\sum_{\ell} {\rm Tr} \left(Z^{\ell}\Phi_{a}Z^{J-\ell} \right)$, 
the value of $(\Delta-J)$ is not corrected at least at first-order in the
weak coupling perturbation theory.
As pointed out by BMN already,  the situation is different for 
operators of the type 
\bea
\sum_{\ell}e^{i\frac{2\pi n \ell}{J}} {\rm Tr} \left(Z^{\ell} (D_{i}Z)
Z^{J-\ell} \right)
\nonumber
\eea
or
\bea
\label{ope2}
\sum_{\ell}e^{i\frac{2\pi n \ell}{J}} {\rm Tr} 
\left(Z^{\ell} \Phi_{i}Z^{J-\ell} \right),
\eea
modulated by the `separation-dependent' phase-factors. 
We will discuss aspects of these corrections in the next subsection.

\subsection{Strings out of Dual Gauge Theory}
Let us first recapitulate what we have done so far. We have considered
Euclidean ${\cal N}=4$ gauge theory around a vacuum state invariant under 
SO(4) $\otimes$ SO(4) and under the outer-automorphism $H$. The Hilbert
space of small fluctuations around this vacua define a representation of
the Heisenberg algebra {\tt h}(8). The outer-automorphism $H$ is simply
the number operator associated with the creation and annihilation operators 
generating {\tt h}(8). In addition, using the gravity
dual of the Minkowskian ${\cal N}=4$ gauge theory, we can define a 
precise map between the creation anhilation operators and
the outer-automorphism $H$ on the dual gauge theory side 
and the Killing vectors in
the Penrose limit of $AdS_{5}\times S^{5}$ on the gravity side. 
The formal conjugate variables
of the Hamiltonian 
$H$ and the central extension $C$ 
become the coordinates $x^{+}$ and $x^{-}$
of the Penrose limit of the bulk spacetime. 

Probably the most surprising result of all this is the connection
between the Penrose limit of Minkowskian ${\cal N}=4$ gauge theory and
the Euclidean ${\cal N}=4$ gauge theory around a particular vacua. The 
main reason for this strange connection has to do
with the peculiarities of spontaneous breakdown of conformal invariance.
In fact, one can in principle think this as a spontaneous breakdown of
SO(4,2) to the Euclidean subgroup SO(4). In this case, the 
Lorentz invariance of the original Minkowskian theory
should be hidden somehow in the dynamics of the Goldstone bosons
around the vacua used to break the conformal symmetry spontaneously.

How then is the Lorentz invariance realized in the Hilbert space of small
fluctuations around the chosen vacua $ \vert 0 \big>_J$?
The answer descending from the BMN proposal is quite surprising and
in fact extremely interesting. It asserts that
the Hilbert space of small fluctuations of the \underline{Euclidean} 
${\cal N}=4$ gauge theory around the vacua $ \vert 0 \big>_J$ 
in the large-N and large-$J$ limit is the Hilbert space of a 
ten-dimensional string theory in a suitable \underline{Minkowskian} background.
 
To understand this, we begin with recalling some salient features of 
string dynamics in the light-cone gauge. In flat ten-dimensional space-time 
and for the bosonic sector, the light-cone gauge-fixed string is defined by: 
\hfill\break
\hfill\break
i) {\tt string oscillators}: infinite tower of Heisenberg algebras
\bea
\left[a_{n}^{i},\, a_{m}^{j \dag} \right] = \delta_{n,m} \delta_{i,j}
\nn
\eea
with $i,j = 1...8$ the transversal coordinates,
\hfill\break
ii) {\tt light-cone Hamiltonian}:
\bea
H_{\rm LC} = \sum_{n} \frac{n}{2 \alpha' p^{+}} a_{n}^{\dag} a_{n} + ({\rm
h.c.}),
\nn
\eea
iii) {\tt string parameter space}: total length of the light-cone string is given by $p^{+}$,
\hfill\break
iv) {\tt Virasoro constraint}: infinite tower
of constraints satisfying the Virasoro algebra. 
\hfill\break
\hfill\break
The way string dynamics emerges out of the Euclidean ${\cal N}=4$ 
gauge theory around the vacuum $\vert 0 \big>_J$ relies crucially on 
the existence of the Heisenberg algebra {\tt h}(8) and of the 
outer-automorphism $H$. In order to establish a connection between the two 
structures, the first thing we should do is to extend the Heisenberg algebra 
{\tt h}(8) to an infinite family of Heisenberg algebras of the type 
displayed in (i). Remarkably, this is achieved by the phase-factor-modulated
operators, Eq.(\ref{ope2}). Introduce creation and annihilation operators
$b_n, b_n^\dagger$ via:
\bea
\sum_{\ell}e^{i \frac{2\pi n \ell}{J}} {\rm Tr} \left(Z^{\ell}
\Phi_{a}Z^{J-\ell} \right)(0) \vert 0 \rangle_{\rm YM} :=
b_{n}^{i \dag} \vert 0 \big>_J.
\nn
\eea
One can show readily that the newly introduced creation and annihilation
operators obey the requisite infinite towers of Heisenberg algebras:
\bea
\label{hei2}
\left[b_{n}^{i}, \, b_{m}^{j \dag} \right] = \delta_{n,m} \delta_{i,j}
\qquad \quad (m, n = 0,1,2, \cdots).
\nn
\eea
Compared to the light-cone string in flat spacetime, the main difference
is the existence of the Heisenberg algebra for the $b_{0}^{i}, \, 
b_0^{i \dag}$ harmonic oscillator operators. In
fact, in flat spacetime, one only has the Heisenberg algebra for the 
center-of-mass part
\bea
\left[x_{0}^{-}, \, p^{+} \right]= i.
\nn
\eea
Let us now see how the outer-automorphism $H$ is modified by quantum
effects.
As computed by BMN, the scaling dimension $\Delta$
for the state $b_{n}^{i \dag} \vert 0 \big>_J$ is given in first-order
in weak-coupling perturbation theory by 
\bea
\Delta = J+1 - {1 \over \pi} g^2_{\rm YM} N 
\left(\cos \left(\frac{2 \pi n}{J} \right)-1 \right).
\nonumber
\eea
Thus, the outer-automorphism at the quantum level is given by 
\bea 
H  =  \left(\Delta -J \right) &=& 1 - {1 \over \pi} g^2_{\rm YM} N 
\left(\cos \left(\frac{2 \pi n}{J} \right)-1 \right)
\nn \\
&=& 1 + \frac{2 \pi g^2_{\rm YM} N}{J^{2}} n^2 + \cdots,
\nn
\eea
where, in obtaining the second expression, we have taken the large-$J$ and
large-$N$ limit. 

The light-cone Hamiltonian in the bulk may be therefore written as
\bea
p_+ = 
\mu \sum_{n} {\sqrt{1+ {4\pi g^2_{\rm YM} N \over J^2}n^2}} \, ~~b_{n}^{\dag}b_{n}
\nonumber
\eea
However, from Eq.({\ref{eq:pminus}), we have
\bea
p_- = {1\over 2\mu R^2}(\Delta + J) \sim {J \over \mu R^2}
= {J \over \mu {\sqrt{4\pi g^2 N}}}
\nonumber
\eea
where we have used $\alpha ' = 1$ units throughout. Thus,
\bea
p_+ = 
\mu \sum_{n} {\sqrt{1 + {n^2 \over \mu^2 p_-^2}}} \, ~~b_{n}^{\dag}b_{n},
\nonumber
\eea
yielding exactly the light-cone Hamiltonian of the string theory
in this background.

An issue we have not elaborated in detail is 
the string Virasoro constraints. We end this section with brief remarks
on it. By inserting phases into the dual gauge theory operators, the 
Heisenberg algebra {\tt h}(8) is extended to the family of Heisenberg 
algebras Eq.(\ref{hei2}). 
It is also natural to extend the outer-automorphism to this
collection of Heisenberg algebras by introducing operators $H_{n}, \,\, (n=0, 1, 2, \cdots)$
such that $\left[H_{n},b_{m} \right] = b_{n+m}$ and $H_{0}=H$ for the 
light-cone Hamiltonian $H$. These outer-automorphisms  then generate  
the string Virasoro algebra. A very interesting question left
for the future would be to uncover meaning of these Virasoro constraints 
entirely within the dual gauge theory viewpoint.

\subsection{Light-Cone Hamiltonian and Renormalization Group Flow}
The renormalization group equation for the correlators $
\Big<\, {\cal O}(x) {\cal O}^{*}(0) \, \Big>$ is
\bea
\left[ a \partial_{a} + 2 \gamma({\cal O}) \right] 
\Big<\, {\cal O} (x) {\cal O}^{*}(0)\, \Big> = 0.
\nn
\eea
where $a$ refers to the renormalization scale.
If we consider the phase-modulated operator ${\cal O} = 
\sum_{n} e^{2 \pi n \ell} {\rm Tr} \left(Z^{\ell}\Phi_a Z^{J-\ell} \right)$,
in the large-$J$ and large-$N$ limit, we get 
\bea
2 \gamma \left({\cal O} \right) = \left(\Delta -J \right)[{\cal O}] -1,
\label{rg}
\eea
where $(\Delta -J)[{\cal O}]$ is the value of $(\Delta -J)$ for the 
operator ${\cal O}$. This equation is re-expressible in a more suggestive
form
\bea
2 \gamma({\cal O} ) {\cal O} = \left[H,{\cal O} \right] - {\cal O},
\nn
\eea
where $H$ is, as usual, the string light-cone Hamiltonian
$(L_{0} + \overline{L}_0)$.

Note that the anomalous dimension $\gamma({\cal O})$ 
appearing in Eq.(\ref{rg}) is, 
for the operator ${\cal O} = \sum_{n} e^{2 \pi i n \ell}
{\rm Tr} \left(Z^{\ell}\Phi_a Z^{J-\ell} \right)$, given by
\bea
\gamma({\cal O}) = J \gamma(Z) + \gamma(\Phi_a),
\nn
\eea
where $\gamma(Z)$ represents the anomalous dimension of the operator $Z$.
Generically, the anomalous dimension $\gamma(Z)$ is affected by radiative 
effects through the self-energy corrections \footnote{These are so-called
zero-momentum effects in the nomenclature of BMN.}.
If not protected by supersymmetry, these contributions would grow as 
$g^2_{\rm YM} N$ and the connection with the string light-cone Hamiltonian 
would be lost. Moreover, the scaling dimension of the operators ${\cal O}$,
in that case, would grow with t'Hooft's coupling constant and eventualy 
disappear out of the physical spectrum. In the 
supersymmetric case we are considering, supersymmetry renders $\gamma(Z)=0$.
This is the reason behind regarding $\gamma({\cal O})$ as the anomalous 
dimension of the field $\Phi_a$'s. From the previous discussion, 
it should be evident that changes in the holographic coordinate 
$x^{+}$ are equivalent to changes of the renormalization scale $a$ 
if we interpret $(\Delta -J)$ as the anomalous dimension \footnote{ 
Definition of the renormalization group $\gamma({\cal O})$ as
$(\Delta -J)$ or as $(\Delta -J) -1$ depends
on whether one adpots the canonical dimensions for the fields and masses
or not. For a lucid discussion on this point, see \cite{'tHooft}}.

\section{Symmetry Enhancement as ``Post Mortem'' Effect}
In recent papers \cite{mukhigomis, orbifolds}, the results of BMN have
been generalized to gauge theories with less supersymmetry, in
particular, to the gravity duals of $AdS_{5} \times T^{1,1}$. The fact
that the Penrose limit of this space is the same as the one of
$AdS_{5} \times S^{5}$ raises the question regarding reason behind
supersymmetry enhancement from ${\cal N}=1$ to ${\cal N}=4$. The point
of this seemingly mysterious result goes back to the fact that, in the
{\sl strict} Penrose limit, there always appear extra isometries.  In
\cite{bfp2}, these extra isometries are referred to as a
\underline{post mortem} effect.  These isometries define always a
Heisenberg algebra. In the case of $AdS_{5} \times T^{1,1}$, the
isometries are SO(4,2)$\otimes$ (SU(2) $\otimes$ SU(2) $\otimes$
U(1)), and are in correspondence, respectively, with the
conformal invariance and the R-symmetry of the dual gauge
theory. Following our approach, we can think in terms of a spontaneous
breakdown of this symmetry to SO(4)$\otimes$SO(4). The difference with
the case of ${\cal N}$=4 supersymmetric gauge theory is that we now
have a smaller number of Goldstone bosons associated with the broken
symmetries, viz. ten instead of eighteen in the ${\cal N} =4$
supersymmetric case.  The deficit eight Goldstone bosons are precisely
the ones associated with the Higgs fields in the ${\cal N}=4$
supersymmetric gauge theory.  These are the fields that would render the
theory ${\cal N}=4$ supersymmetric.  In the Penrose limit of any
ten-dimensional background, we have always a Heisenberg algebra {\tt
h}(8) of the isometries.  In the $AdS_{5} \times T^{1,1}$ case, the
Heisenberg algebra is composed of two Heisenberg subalgebras {\tt
h}(4) with a common central extension.  The eight generators of one of
the two algebras are the ones we are going to use as the eight missing
broken symmetries. These ``post mortem'' Goldstone bosons are simply
states that, in the ${\cal N}=1$ gauge theory, are degenerate in the
light-cone mass (the eigenvalue of the outer-automorphism or the light-cone
Hamiltonian ) with the real Goldstone bosons. As the enhancement of
${\cal N}=4$ supersymmetry is true only in the strict Penrose limit,
viz. for the Penrose scaling factor $R \rightarrow \infty$, we expect 
this enhancement to be violated by ${\cal O}(1/N)$-corrections.

\section{Conclusions}
Our main conclusion is that the supersymmetric gauge theory dual to the
IIB string theory in a ten dimensional pp-wave background 
may be thought to ``live'' on a 
\underline{Euclidean}
four dimensional space. The indications has come from several corners. 
The most
direct reason is that the dual operators are in one to one correspondence
with the states of the string theory which are created by {\it transverse}
oscillators in the light-cone gauge. This is in fact apparent in the 
zero-mode sector of the string relevant for the supergravity modes and 
follows from field equations in this background. Furthermore,
understanding of the low-energy Nambu-Goldstone modes resulting 
from symmetry breaking of the original symmetry group SO(4,2)$\times$SO(6) to 
SO(4)$\times$SO(4)$\times$H(4)$\times$H(4) 
also indicates that the dual theory lives in a Euclidean space.

One may think of this Euclidean space as the space spanned by four of the
transverse directions. In this case, the light-cone time has a natural
explanation as the holographic coordinate representing the scale of 
the dual gauge theory. Details of the proposed holographic correspondence 
remain to be understood better. 
The fact that a Euclidean theory can give rise to
strings living in a space-time with Lorentzian signature is intruiging and 
deserves a better understanding. 
We expect that this would enhance our understanding of holography in general.

Extending the results of this work, one can abstract the main ingredients
needed in order to generalize the BMN proposal to generic situations. 
The minimal starting point would be a four-dimensional gauge theory
invariant under the conformal group and with a nonanomalous
global U(1) symmetry. The spontaneous breakdown of
SO(4,2)$\otimes$U(1)  to SO(4) provides ten broken generators that we can 
try to organize, taking Euclidean signature dual theory,
into a Heisenberg algebra {\tt h}(4) and an outer-automorphism $H$. 
This is the structure that we can try to map into the Hilbert space of a
non-critical six-dimensional string theory
in the light-cone gauge with the SO(4) rotational invariance acting on
the transversal coordinates. Of course, we also need controllable, finite 
contributions- in the large 't Hooft coupling limit - to the anomalous 
dimensions of the fields representing the small quantum fluctuations around 
the selected vacuum or, equivalently,
finite light-cone mass (defined by the outer-automorphism ) for the 
Goldstone bosons. Taking the real-world QCD, one now has
{\sl ab initio} two 
related problems: anomaly for the conformal invariance (viz. a
non vanishing beta function) and anomaly for the axial U(1) symmetry,
which would serve as a natural candidate for the global U(1) symmetry. 
We can try to solve these problems by introducing two extra scalar fields, 
namely the dilaton $D$ and the axion $A$. We then find an indication that
the field $(D+iA)$ is a natural candidate to play the role of the $Z$-field 
in the BMN proposal. 

\section*{Acknowledgements}
We would like to thank the Isaac Newton Institute for Mathematical Sciences,
Cambridge University and organizers of `M-theory' workshop for
hospitality. The work of S.R.D. is partially supported by U.S. DOE
contract DE-FG01-00ER45832.  The work of C.G is partially supported
by grant AEN2000-1584.

\subsection*{Note Added:}
After this paper was submitted in the preprint archive, two preprints
\cite{kp}, \cite{lor} investigating related issues have appeared. The
paper \cite{kp} argues that the hologram is eight dimensional, based
on the classical Cauchy problem - rather than four dimensional.  In
\cite{lor} the holographic screen is argued to be four-dimensional
Minkowski space, but the realization of the SO(4) symmetry from this
viewpoint was not addressed.

\end{document}